\documentclass[aps,prd,preprint,showpacs,floatfix,nofootinbib]{revtex4-1}

\usepackage{dcolumn}

\usepackage{amsmath, amsfonts, amssymb, mathrsfs, times, float, color, bm}
\usepackage{extarrows}
\usepackage{graphicx}
\usepackage{graphicx,epstopdf}
\usepackage{dcolumn}
\usepackage{bm}

\usepackage{exscale}
\usepackage{relsize}

\newcommand{\nn}{\nonumber}

\begin{document}

\title[]{Second-order time delay by a radially moving Kerr-Newman black hole}

\author{Guansheng He}
\author{Wenbin Lin}
\email{wl@swjtu.edu.cn}

\affiliation{School of Physical Science and Technology, Southwest Jiaotong University, Chengdu 610031, People's Republic of China}

\begin{abstract}
We derive the analytical time delay of light propagating in the equatorial plane and parallel to the velocity of a moving Kerr-Newman black hole up to the second post-Minkowskian order via integrating the null geodesic equations. The velocity effects are expressed by a very compact form. We then concentrate on analyzing the magnitudes of the correctional effects on the second-order contributions to the delay and discuss their possible detection. Our result in the first post-Minkowskian approximation is in agreement with Kopeikin and Sch\"{a}fer's formulation which is based on the retarded Li\'{e}nard-Wiechert potential.

\begin{description}
\item[PACS numbers] 98.62.Sb, 95.30.Sf, 04.70.Bw, 04.25.Nx
\end{description}

\end{abstract}

\maketitle

\section{Introduction}

The time dependence of a background field caused by the translational motion of a gravitational source usually exerts an influence on the propagation of electromagnetic waves. This kind of kinematical effect (also called the velocity effect~\cite{Heyrovsky2005}) on the gravitational time delay of light has been investigated in detail in the last two decades~\cite{KopeiSch1999,KopeiMash2002,Sereno2002,Frittelli2003,Sereno2005,KopeiFoma2007,BAI2008,Kopeikin2009,HBL2014,SH2015}. In particular, Kopeikin and Sch\"{a}fer~\cite{KopeiSch1999} pioneered the Lorentz-covariant theory for light propagating in the gravitational field of an ensemble of arbitrarily moving bodies, in which the generalized form of the Shapiro time delay~\cite{Shapiro1964,Shapiro1971} was obtained in the first post-Minkowskian (1PM) approximation. Their calculations were based on the Li\'{e}nard-Wiechert gravitational potential and later extended in Ref.~\cite{KopeiMash2002} to investigate the spin-dependent gravitomagnetic effects. Sereno~\cite{Sereno2002,Sereno2005} employed Fermat's principle~\cite{SEF1992} to study the gravitational lensing caused by a slowly moving, spinning body in the framework of the standard lens theory, including the kinematically correctional effects on the light delay. Recently, the time transfer functions proposed by Teyssandier and Le Poncin-Lafitte~\cite{TL2008} were applied to calculate the observable relativistic effects containing the Shapiro effect in the field of moving axisymmetric bodies~\cite{HBL2014}. This approach was also used to deal with the light delay due to a moving gravitational source with a low velocity and arbitrary multipoles~\cite{SH2015}.

These previous surveys of the velocity corrections were mainly aimed at the first-order gravitational signals delay. The magnitudes of these correctional effects are so relatively large that they are very likely to be detected, whether the motion of the gravitational source is relativistic or not. Not to be forgotten, the nonrelativistic velocity effects (appearing as extrinsic gravitomagnetic effects~\cite{KopeiFoma2007,CW1995}) were confirmed by the Jovian deflection experiment in 2002~\cite{Frittelli2003,FK2003,KopeiFoma2007}, with an accuracy of $20\%$. As is known, nowadays techniques in the Shapiro delay measurements have made rapid progresses and achieved a high precision at the picosecond $(ps\sim 10^{-12}s)$ level or even better. For example, the delay precision of the next generation of the Very-Long-Baseline Interferometry (VLBI) system was proposed to be $4ps$~\cite{Petrachenko2009,SB2012}. One can expect that the kinematical effects on the second-order time delay might also be detectable and, therefore, deserve our attentions. It requires full theoretical treatment of the gravitational retardation effect induced by a moving lens in the second post-Minkowskian (2PM) approximation.

In the present paper, we investigate the velocity effects on the second-order gravitational delay of light propagating in the equatorial plane of a moving Kerr-Newman (KN) black hole with a constant radial velocity, which serves as a natural extension of our previous result~\cite{LinHe2016}. We restrict our discussions to the weak-field, small-angle, and thin-lens approximation. For the convenience of the computations, we will define the impact parameter by assuming that light signals come from infinity with an initial velocity being parallel to the black hole's velocity.

This paper is organized as follows. In Sec.~\ref{derivation}, we first review the weak-field metric of the moving KN source, and then derive the explicit time delay up to the 2PM order. Section~\ref{Magnitude} is devoted to estimating the magnitudes of the velocity effects on the delay for three typical cases of the lens' mass. In Sec.~\ref{detection}, we discuss the possibility of detecting the correctional effects on the second-order contributions to the delay. The summary is given in Sec.~\ref{Summary}. In what follows, we use natural units in which $G = c = 1$.

\section{Second-order moving Kerr-Newman time delay} \label{derivation}

\subsection{The 2PM metric for a moving KN source with a constant radial velocity}

The second post-Minkowskian metric of a radially moving Kerr-Newman black hole can be obtained from the harmonic Kerr-Newman metric~\cite{LinJiang2014} via a Lorentz boost transformation. We assume $\{\bm{e}_1,~\bm{e}_2,~\bm{e}_3\}$ to be the orthonormal basis of a three-dimensional Cartesian coordinate system. Let $(t,~x,~y,~z)$ and $(X_0,~X_1,~X_2,~X_3)$ denote the rest frame of the background and the comoving frame for the barycenter of the gravitational source, respectively. The 2PM harmonic metric of a moving KN black hole with a constant radial velocity $\bm{v}=v\bm{e_1}$ can be written as~\cite{LinHe2015}
\begin{eqnarray}
&&g_{00}=-1\!+\!\frac{2(1\!+\!v^2)\gamma^2M}{R}\!-\!\frac{M^2\!+\!\gamma^2(M^2\!+\!Q^2)}{R^2}\!-\!\frac{4v\gamma^2aM X_2}{R^3}
\!+\!\frac{v^2\gamma^2(M^2\!-\!Q^2)X_1^2}{R^4}\!+\!O(G^3)~,~~~~ \label{g00MKN} \\
\nn&&g_{0i}=\gamma\zeta_i-v\gamma^2\!\left(\frac{4M}{R}\!-\!\frac{M^2\!+\!Q^2}{R^2}\right)\!\delta_{i1}
\!-\!\frac{v\gamma\,(M^2\!-\!Q^2)X_1\left[X_i\!+\!(\gamma\!-\!1)X_1\delta_{i1}\right]}{R^4}\!+(\gamma^2\!-\!\gamma\!+\!v^2\gamma^2) \!\times   \\
&& \hspace*{29pt} \frac{2\,aM X_2\,\delta_{i1}}{R^3}+O(G^3)~,  \label{g0iMKN} \\
\nn&&g_{ij}=\left(1\!+\!\frac{M}{R}\right)^{\hspace*{-1.5pt}2}\delta_{ij}
\!+\!v^2\gamma^2\left(\frac{4M}{R}\!-\!\frac{M^2\!+\!Q^2}{R^2}\right)\delta_{i1}\delta_{j1}
\!-\!v\,\gamma \!\left[\zeta_i\,\delta_{j1}\!+\!\zeta_j\,\delta_{i1}\!+\!\frac{4(\gamma \!-\!1)aM X_2\,\delta_{i1}\,\delta_{j1}}{R^3} \right] \\
&& \hspace*{29pt}+\frac{(M^2-Q^2)\left[X_i+(\gamma-1)X_1\delta_{i1}\right]\left[X_j+(\gamma-1)X_1\delta_{j1}\right]}{R^4}+O(G^3)~,  \label{gijMKN}
\end{eqnarray}
where $i$ and $j$ take values among the set $\{1,~2,~3\}$, $\delta_{ij}$ denotes the Kronecker delta, and $\gamma=(1-v^2)^{-\scriptstyle \frac{1}{2}}$ is the Lorentz factor. $M$, $Q$, and $\bm{J} (=\!J\bm{e}_3)$ are the rest mass, electrical charge, and angular momentum vector of the gravitational source, respectively. $\Phi\equiv-\frac{M}{R}$ represents Newtonian gravitational potential, with $\frac{X_1^2+X_2^2}{R^2+a^2}\!+\!\frac{X_3^2}{R^2}\!=\!1$ and $\mathbf{X} \! \cdot \! d\mathbf{X}\equiv X_1dX_1 \!+\! X_2dX_2 \!+\! X_3dX_3$. $a \!\equiv\! \frac{J}{M}$ is the angular momentum per mass, and $\bm{\zeta}\!\equiv\!\frac{2aM}{R^3}\left(\bm{X} \!\times\! \bm{e_3}\right)=(\zeta_1,~\zeta_2,~0)$. The relation $M^2\geq a^2+Q^2$ is assumed to avoid naked singularity for the black hole. Notice that the coordinates $X_0,~X_1,~X_2$, and $X_3$ are related to $t,~x,~y$, and $z$ by the Lorentz transformation as follow
\begin{eqnarray}
&& X_0=\gamma(t-vx)~, \label{LT1} \\
&& X_1=\gamma(x-vt)~, \label{LT2} \\
&& X_2=y~,            \label{LT3} \\
&& X_3=z~.            \label{LT4}
\end{eqnarray}

\subsection{Second-order time delay caused by the moving KN black hole} \label{TimeDelay}
\begin{figure*}[t]
\begin{center}
  \includegraphics[width=14cm]{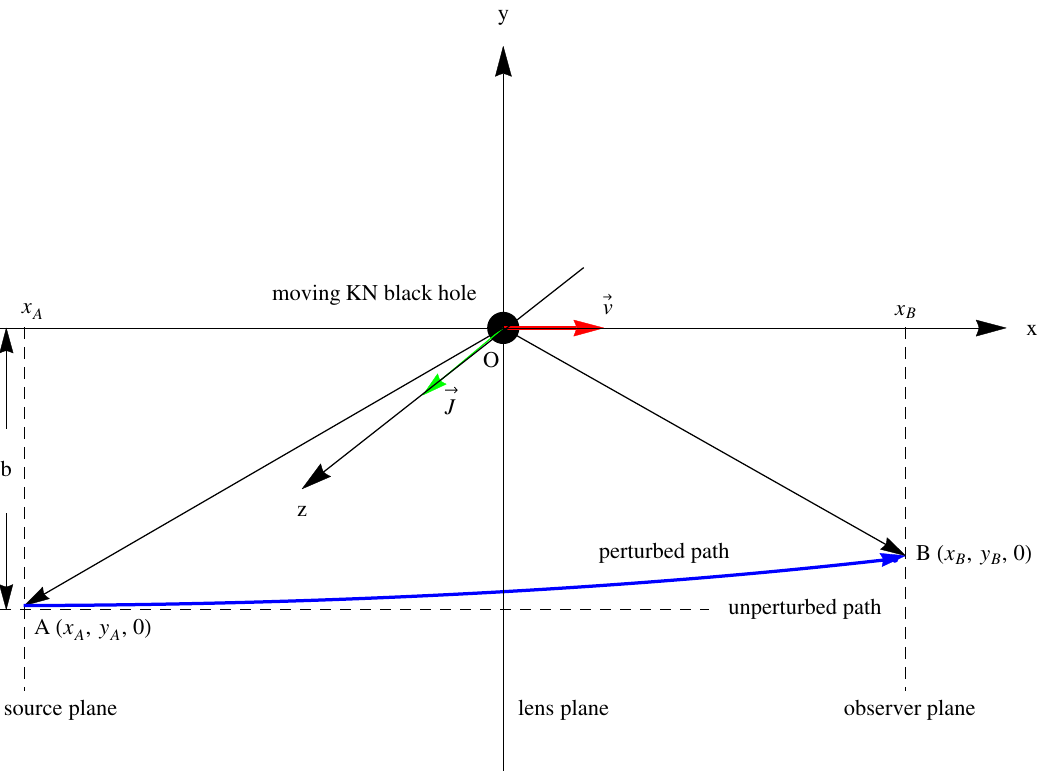}
  \caption{Schematic model for light propagating in the gravitational field of the moving KN black hole. The gravitational deflection is greatly exaggerated to distinguish the blue line (perturbed path) from the dashed horizontal line (unperturbed path). Light is supposed to take the prograde motion relative to the rotation $\bm{J}$ of the gravitational source.}    \label{Figure1}
\end{center}
\end{figure*}
We consider the gravitational time delay of light caused by the moving Kerr-Newman black hole. For simplicity, light signals are assumed to propagate in the equatorial plane of the black hole $(z=\frac{\partial }{\partial z}=0)$. In contrast to the gravitational deflection case where the light emitter and receiver can be set at infinity, for the time delay, the emitter and receiver cannot be located at infinity, otherwise the time delay will become infinity. Hence, we set the emitter and receiver to be located at the points $A$ and $B$, respectively, both of which are far away from the lens.

The schematic model for light propagation is shown in Fig.~\ref{Figure1}. The spatial coordinates of the light emitter (denoted by $A$ on the source plane) and receiver (denoted by $B$ on the observer plane) are assumed to be $(x_A,~y_A,~0)$ and $(x_B,~y_B,~0)$, respectively, in the background's rest frame, where $y_A<0$, $x_A<0$ and $x_B>0$. $b$ denotes the impact parameter which is strictly defined as follows. Let the blue line represent the propagation path of a photon coming from $p=-\infty$ with the initial velocity $\bm{w}|_{p\rightarrow-\infty}~(=\bm{e}_1)$ being parallel to the central mass's velocity $\bm{v}$, where $p$ denotes the affine parameter of the trajectory~\cite{WuckSperh2004,Weinberg1972}. Then, the impact parameter is defined via the $y$ coordinate of a photon as $b\equiv-y|_{p\rightarrow-\infty}$, which denotes the geometrical distance between the $x$ axis and unperturbed path of light. This definition is convenient for the Cartesian coordinate system, and it is a little bit different from the definition by the conservation of the angular momentum of a photon~\cite{Teyssandier2012} since the black hole is not static. Notice that the locations of $A$ and $B$ are denoted by $(X_A,~Y_A,~0)$ and $(X_B,~Y_B,~0)$ in the comoving frame, respectively.

The general form of the null curve is given in the background's rest frame as
\begin{equation}
0=ds^2=g_{\mu\nu}dx^\mu dx^\nu~,    \label{geodesic1}
\end{equation}
where the indices $\mu$ and $\nu$ run over the values $0,~1,~2,~3$. For light propagating in the equatorial plane, Eq.~\eqref{geodesic1} is reduced to
\begin{equation}
0=g_{00}\,dt^2+g_{11}\,dx^2+g_{22}\,dy^2+2\,g_{01}\,dtdx+2\,g_{02}\,dtdy+2\,g_{12}\,dxdy ~,    \label{geodesic2}
\end{equation}
which results in
\begin{equation}
\frac{dt}{dx}=\frac{-M+\sqrt{M^2-4N}}{2} ~,    \label{geodesic4}
\end{equation}
with
\begin{eqnarray}
&&M=2\left(\frac{g_{01}}{g_{00}}+\frac{g_{02}}{g_{00}}\frac{dy}{dx}\right) ~,  \label{E} \\
&&N=\frac{g_{11}}{g_{00}}+\frac{2g_{12}}{g_{00}}\frac{dy}{dx}+\frac{g_{22}}{g_{00}}\left(\frac{dy}{dx}\right)^2 ~.  \label{F}
\end{eqnarray}
Here, we have ignored the other solution for its nonphysical property, since $N=-1+O(G)\!<0$ in the weak-field and small-angle approximation. Note that $\frac{dy}{dx}$ is related to the gravitational deflection angle $\alpha$ of light by $\left.\alpha=\arctan{\frac{dy}{dx}}\right|_{p\rightarrow -\infty}^{p\rightarrow +\infty}=\left.\arctan{\frac{dy}{dx}}\right|_{x\rightarrow -\infty}^{x\rightarrow +\infty}$.

We then perform an indefinite integral over $x$ for Eq.~\eqref{geodesic4} and get
\begin{equation}
t=\mathlarger{\int}\left[\sqrt{-\frac{g_{11}}{g_{00}}+g_{01}^2+\left(\frac{dy}{dx}\right)^2}-\frac{g_{01}}{g_{00}}\right]dx~,  \label{Integral-1}
\end{equation}
where the third- and higher-order terms (e.g., $\frac{g_{02}}{g_{00}}\frac{dy}{dx}$) have been neglected for the calculations of time delay up to the second order. Notice that the terms with the factor $\frac{aM}{R^2}$ or $\frac{Q^2}{R^2}~\left(\leq \frac{M^2}{R^2}\right)$ in Eqs.~\eqref{g00MKN} - \eqref{gijMKN} are regarded as second-order terms because of the assumption $M^2\geq a^2+Q^2$~\cite{LinJiang2014,ERT2002,EG2006}. This is a little bit different from the definition of the second-order terms in Ref.~\cite{PLT2004}, where these terms are regarded as first-order terms which are of order $O(G)$.

In order to obtain the analytical coordinate time $t$, we only need to calculate the explicit form of $\frac{dy}{dx}$ to the first order. We begin with the 1PM equations of motion for light in the gravitational field of a moving Schwarzschild black hole with the radial velocity $v$~\cite{LinHe2016b}
\begin{eqnarray}
&&0=\ddot{t}+\frac{\left[v(v^2-3)\dot{x}^2+2(1+v^2)\dot{t}\dot{x}-v(1+v^2)\dot{t}^2\right]\gamma^3 M X_1}{R^3}+O(G^2)  ~,   \label{motonEQ-t}   \\
&&0=\ddot{x}+\frac{\left[(1-3v^2)\dot{t}^2+2v(1+v^2)\dot{t}\dot{x}-(1+v^2)\dot{x}^2\right]\gamma^3 M X_1}{R^3}+O(G^2)  ~,  \label{motonEQ-x}   \\
&&0=\ddot{y}+\frac{\left[(1+v^2)\dot{t}^2-4v\dot{t}\dot{x}+(1+v^2)\dot{x}^2\right]\gamma^2MX_2}{R^3}+O(G^2)  ~,  \label{motonEQ-y}
\end{eqnarray}
where a dot denotes the derivative with respect to $p$ which is assumed to be $x$ to calculate the first-order form of $\frac{dy}{dx}$, as done in Ref.~\cite{WuckSperh2004}.

With the help of the boundary conditions $\dot{t}|_{x\rightarrow-\infty}\!=\!1$ and $\dot{x}|_{x\rightarrow-\infty}\!=\!1$, we can obtain the zero-order values for $\dot{t}$ and $\dot{x}$ from Eqs.~\eqref{motonEQ-t} and \eqref{motonEQ-x} as follows:
\begin{eqnarray}
&&\dot{t}=1+O(G)  ~,  \label{motonEQ-t-2}   \\
&&\dot{x}=1+O(G)  ~.  \label{motonEQ-x-2}
\end{eqnarray}
We then substitute Eqs.~\eqref{motonEQ-t-2} and \eqref{motonEQ-x-2} into Eq.~\eqref{motonEQ-y}, and integrate Eq. \eqref{motonEQ-y} over $x$
to get the analytical form for $\dot{y}$, 
\begin{equation}
\frac{dy}{dx}=\frac{2(1-v)\gamma M}{b}\left(1+\frac{X_1}{\sqrt{X_1^2+b^2}}\right)+O(G^2) ~,    \label{motonEQ-y-2}
\end{equation}
where the zero-order parameter transformation $dX_1\!=\!\gamma(dx-vdt)\!=\!(1-v)\gamma dx$~\cite{WuckSperh2004} and the boundary conditions (in the comoving frame) $\dot{y}|_{X_1\rightarrow-\infty}\!=\!0$ and $y|_{X_1\rightarrow-\infty}\!=\!-b$ have been used. Notice that in the limit $X_1\rightarrow +\infty$, the first-order moving Schwarzschild deflection angle can be obtained from Eq.~\eqref{motonEQ-y-2} as $\alpha=\frac{4(1-v)\gamma M}{b}$~\cite{WuckSperh2004}.

Plugging Eqs.~\eqref{g00MKN} and \eqref{gijMKN} and \eqref{motonEQ-y-2} into Eq.~\eqref{Integral-1}, we have
\begin{eqnarray}
\nn&& t=\mathlarger{\int}\Bigg{\{}\,\Bigg{[}\,\frac{1+\frac{2\,(\,1\,+\,v^2\,)\,\gamma^2\,M}{R}+\frac{M^2\,-\,v^2\,\gamma^2\,(\,M^2\,+\,\,Q^2\,)}{R^2}
-\frac{4\,v\,\gamma^2\,a\,M\,y}{R^3}+\frac{\gamma^2\,(\,M^2\,-\,Q^2\,)\,X_1^2}{R^4}}
{1-\frac{2\,(\,1\,+\,v^2\,)\,\gamma^2\,M}{R}+\frac{M^2\,+\,\gamma^2\,(\,M^2\,+\,\,Q^2\,)}{R^2}
+\frac{4\,v\,\gamma^2\,a\,M\,y}{R^3}-\frac{v^2\,\gamma^2\,(\,M^2\,-\,Q^2\,)\,X_1^2}{R^4}}+\frac{16\,v^2\,\gamma^4\,M^2}{R^2}   \\
\nn&& \hspace*{15pt}+\frac{4(1\!-\!v)^2\gamma^2M^2}{b^2}\!\left(\!1\!+\!\frac{X_1}{\sqrt{X_1^2\!+b^2}}\!\right)^{\!\!2} \Bigg{]}^{\!\scriptstyle \frac{1}{2}}
\!\!-\!\frac{v\gamma^2\!\left[\frac{4M}{R}\!-\!\frac{M^2+\,Q^2}{R^2}\!+\!\frac{(M^2-\,Q^2)\,X_1^2}{R^4}\right]
\!-\!\frac{2\,(1+v^2)\,\gamma^2\,aMy}{R^3} }{1-\frac{2(1+v^2)\gamma^2M}{R}} \Bigg{\}}dx¡¡  \\
\nn&&\hspace*{7pt}=\mathlarger{\int} \Bigg{\{} \Bigg{[} 1\!+\!\frac{4(1\!+\!v^2)\gamma^2M}{R}
\!+\!\frac{8(1\!+\!v^2)^2\,\gamma^4M^2\!+\!16v^2\gamma^4M^2\!-\!(1\!+\!v^2)\gamma^2(M^2\!+\!Q^2)}{R^2}\!-\!\frac{8v\gamma^2aMy}{R^3}   \\
\nn&&\hspace*{15pt}+\frac{(1+v^2)\gamma^2(M^2-Q^2)X_1^2}{R^4}+\frac{4\,(1-v)^2\gamma^2M^2}{b^2}\!\left(1+\frac{X_1}{\sqrt{X_1^2+b^2}}\right)^2 \Bigg{]}^{\frac{1}{2}}
\!\!-\frac{v\gamma^2(M^2-Q^2)X_1^2}{R^4}      \\
\nn&&\hspace*{15pt}-\,\frac{4\,v\,\gamma^2M}{R}-\frac{8\,v\,(1+v^2)\,\gamma^4\,M^2-v\,\gamma^2(M^2+Q^2)}{R^2}+\frac{2\,(1+v^2)\,\gamma^2\,a\,My}{R^3} \Bigg{\}}dx \\
\nn&&\hspace*{7pt}=\mathlarger{\int} \! \Bigg{\{}  1+\frac{2(1\!-\!v)^2\gamma^2M}{R}
+\frac{(1\!-\!v)\left[\,(3-5\,v)M^2-(1\!+\!v)\,Q^2\,\right]}{2(1+v)^2R^2}+\frac{(1-v)^2\gamma^2\,(M^2-Q^2)X_1^2}{2R^4}    \\
&&\hspace*{15pt}+\frac{2(1-v)^2\gamma^2M^2}{b^2}\left(\!1\!+\!\frac{X_1}{\sqrt{X_1^2+b^2}}\!\right)^2+\frac{2\,(\,1-v\,)^2\,\gamma^2\,a\,M\,y}{R^3} \Bigg{\}}dx~,~~~~\label{Integral-2}
\end{eqnarray}
where the third- and higher-order terms have been ignored.

To get the explicit form of $y$ up to the 1PM order on the right-hand side of Eq.~\eqref{Integral-2}, we integrate Eq.~\eqref{motonEQ-y-2} over $x$,
\begin{equation}
y=-b\left[1-\frac{2M\left(\sqrt{X_1^2+b^2}+X_1\right)}{b^2}+O(G^2)\right]~,~~\left(\frac{2M(\sqrt{X_1^2+b^2}+X_1)}{b^2}\ll1\right)~,~ \label{y}
\end{equation}
where the zero-order approximation $dX_1=(1-v)\gamma dx$ and the boundary condition $\left.y\right|_{X_1\rightarrow -\infty}=-b$ have been used. Substituting Eq.~\eqref{y} into Eq.~\eqref{Integral-2}, we can obtain
\begin{eqnarray}
\nn&&t=\! \mathlarger{\int}\! \Bigg{\{}  1\!+\!\frac{2\,(1\!-\!v)^2\,\gamma^2M}
{\sqrt{X_1^2\!+\!b^2}\sqrt{1\!-\!4M\left(\!\sqrt{X_1^2\!+\!b^2}\!+\!X_1\!\right)\!/\!\left(X_1^2\!+\!b^2\right)}}
\!+\!\frac{(1\!-\!v)\left[(3\!-\!5\,v)M^2\!-\!(1\!+\!v)Q^2\right]}{2\,(\,1+v\,)^2\left(X_1^2+b^2\right)}    \\
\nn&&\hspace*{18pt}+\frac{2(1\!-\!v)^2\gamma^2M^2}{b^2}\!\left(\!1\!+\!\frac{X_1}{\sqrt{X_1^2\!+\!b^2}}\!\right)^2
\!+\!\frac{(1\!-\!v)^2\gamma^2(M^2\!-\!Q^2)X_1^2}{2\left(X_1^2+b^2\right)^2}\!+\!\frac{2(1\!-\!v)^2\gamma^2aMy}{\left(X_1^2+b^2\right)^{ \frac{3}{2}}} \Bigg{\}}dx   \\
\nn&&\hspace*{8pt} = x+(1-v)^2 \gamma^2\! \mathlarger{\int} \!\Bigg{[}\frac{2\,M}{\sqrt{X_1^2+b^2}}
\!+\!\frac{4\,M^2\!\left(\!\sqrt{X_1^2+b^2}\!+\!X_1\!\right)}{\left(X_1^2+b^2\right)^{\frac{3}{2}}}\!+\!\frac{\frac{(3-5\,v)\,M^2}{1+v}\!-\!Q^2}{2\left(X_1^2+b^2\right)}
\!+\!\frac{(M^2\!-\!Q^2)X_1^2}{2\left(X_1^2+b^2\right)^2}    \\
&&\hspace*{18pt}+\,\frac{2M^2}{b^2}\left(1+\frac{X_1}{\sqrt{X_1^2+b^2}}\right)^{\hspace*{-2pt}2}-\frac{2aMb}{\left(X_1^2+b^2\right)^{\frac{3}{2}}} \Bigg{]}dx~, \label{Integral-3}
\end{eqnarray}
with the third- and higher-order terms being dropped in the derivation.

In order to integrate the second part on the right-hand side of Eq.~\eqref{Integral-3} more conveniently, we perform a coordinate transformation
$dX_1=\gamma(1- v\dot{t}/\dot{x})dx$ which is to be calculated up to the first post-Minkowskian order. We substitute Eqs.~\eqref{motonEQ-t-2} and \eqref{motonEQ-x-2} into Eqs.~\eqref{motonEQ-t} and \eqref{motonEQ-x}, and integrate the latter over $x$ to obtain the explicit forms of $\dot{t}$ and $\dot{x}$ up to the 1PM order as follows:
\begin{eqnarray}
&&\dot{t}=1+\frac{2(1-v)\gamma^2 M}{\sqrt{X_1^2+b^2}}+O(G^2) ~,   \label{dott}   \\
&&\dot{x}=1+\frac{2v(1-v)\gamma^2 M}{\sqrt{X_1^2+b^2}}+O(G^2) ~.   \label{dotx}
\end{eqnarray}
Thus, we can get
\begin{eqnarray}
\nn&&dX_1=\gamma\left[1-v\left(1+\frac{2(1-v)^2\gamma^2M}{\sqrt{X_1^2+b^2}}\right)+O(G^2)\right]dx    \\
&&\hspace*{24pt}=(1-v)\gamma\left[1-\frac{2v(1-v)\gamma^2M}{\sqrt{X_1^2+b^2}}+O(G^2)\right]dx~.  \label{transform}
\end{eqnarray}
Notice that up to the zero order Eq.~\eqref{transform} reduces to $dX_1=(1-v)\gamma dx$, which is enough to deal with the 1PM gravitational deflection~\cite{WuckSperh2004}.
Plugging Eq.~\eqref{transform} into Eq.~\eqref{Integral-3}, we have
\begin{eqnarray}
\nn&&t= x+(1\!-\!v)\gamma\!\mathlarger{\int}\!\Bigg{[} \frac{2 M}{\sqrt{X_1^2\!+\!b^2}}+\frac{2 M^2}{b^2}\!\left(\!1\!+\!\frac{X_1}{\sqrt{X_1^2\!+\!b^2}}\!\right)^{\hspace*{-2pt}2}
\!+\!\frac{3M^2-Q^2}{2\left(X_1^2\!+\!b^2\right)}+\frac{4 M^2\left(\!\sqrt{X_1^2\!+\!b^2}\!+\!X_1\!\right)}{\left(X_1^2+b^2\right)^{\frac{3}{2}}}    \\
\nn&&\hspace*{18pt}+\,\frac{(M^2-Q^2)X_1^2}{2\left(X_1^2+b^2\right)^2}-\frac{2aMb}{\left(X_1^2+b^2\right)^{\frac{3}{2}}} \Bigg{]}dX_1   \\
\nn&&\hspace*{8pt}=x+(1-v)\gamma\Bigg{[}2M\,ln\!\left(\!\!\sqrt{\!X_1^2\!+\!b^2}\!+\!X_1\!\!\right)\!+\!\frac{4M^2\!\left(\!\sqrt{\!X_1^2\!+\!b^2}\!+\!X_1\!\right)}{b^2}
\!+\!\frac{3(5M^2-Q^2)}{4b}\,ArcTan\frac{X_1}{b}   \\
&&\hspace*{18pt}-\frac{4M^2}{\sqrt{X_1^2+b^2}}-\frac{(M^2-Q^2)X_1}{4\left(X_1^2+b^2\right)}-\frac{2aM X_1}{b\sqrt{X_1^2+b^2}}\Bigg{]}+C ~, \label{Integral-4}
\end{eqnarray}
where $C$ denotes the integral constant and the third- and higher-order terms have been dropped.

Finally, the explicit form of the time delay up to the second order for light propagating from the light emitter $A$ to the receiver $B$ can be calculated from  Eq.~\eqref{Integral-4} as follows:
\begin{eqnarray}
\nn&&t\,(B,~A)=(x_B-x_A)\!+\!(1-v)\gamma \! \left[2M\,ln\left(\!\frac{\sqrt{X_B^2\!+\!b^2}+X_B}{\sqrt{X_A^2\!+\!b^2}+X_A}\!\right)
\!+\!4M^2\!\left(\!\frac{1}{\sqrt{X_A^2\!+\!b^2}}\!-\!\frac{1}{\sqrt{X_B^2\!+\!b^2}}\!\right)  \right.  \\
\nn&&\hspace*{40pt}\left. +\frac{4M^2}{b^2}\!\left(X_B-X_A+\sqrt{X_B^2+b^2}-\sqrt{X_A^2+b^2}\,\right)
+\frac{M^2-Q^2}{4}\!\left(\frac{X_A}{X_A^2+b^2}-\frac{X_B}{X_B^2+b^2}\right) \right.  \\
&&\hspace*{40pt}\left. +\frac{15M^2\!-\!3\,Q^2}{4\,b}\!\left(\!\arctan{\frac{X_B}{b}}\!-\!\arctan{\frac{X_A}{b}}\!\right)
\!+\!\frac{2\,aM}{b}\!\left(\!\frac{X_A}{\sqrt{X_A^2\!+\!b^2}}\!-\!\frac{X_B}{\sqrt{X_B^2\!+\!b^2}}\!\right) \right]~,      \label{Integral-5}
\end{eqnarray}
where $X_A=\gamma (x_A-v t_A)$ and $X_B=\gamma (x_B-v t_B)$.

\subsection{Discussion of the result} \label{Discussions}
The first term on the right-hand side of Eq.~\eqref{Integral-5}, which is independent of the gravitational source, represents the geometrical time for light traveling in a straight line. In the first post-Minkowskian approximation, Eq.~\eqref{Integral-5} reduces to
\begin{equation}
t\,(B,~A)=(x_B-x_A)+2(1-v)\gamma M \,ln \left(\frac{\sqrt{X_B^2+b^2}+X_B}{\sqrt{X_A^2+b^2}+X_A}\right)~. ~~~\label{Integral-6}
\end{equation}

For a nonmoving Kerr-Newman source, Eq.~\eqref{Integral-5} can be simplified to~\cite{LinHe2016}
\begin{equation}
t\,(B,~A)=(x_B-x_A)\!+\! 2M \, ln\!\left(\frac{\sqrt{x_B^2+b^2}+x_B}{\sqrt{x_A^2+b^2}+x_A}\right)
\!+\!\frac{8M^2x_B}{b^2}\!+\!\frac{15\pi M^2}{4b}\!-\!\frac{4\,aM}{b}\!-\!\frac{3\pi Q^2}{4b}~, \label{Integral-7}
\end{equation}
where the assumptions $x_A\ll-b$ and $x_B\gg b$ have been considered, and the second-order terms with the factor $\frac{1}{x_A}$ or $\frac{1}{x_B}$ have been dropped since they are found to be much smaller than those containing the factor $\frac{1}{b}$.

Provided the angular momentum and electrical charge are dropped from the gravitational source $(a=Q=0)$, Eq.~\eqref{Integral-5} is reduced to the second-order moving Schwarzschild time delay, which reads
\begin{eqnarray}
\nn&&t\,(B,~A)=(x_B-x_A)\!+\!(1-v)\gamma \! \left[2M\,ln\left(\!\frac{\sqrt{X_B^2\!+\!b^2}+X_B}{\sqrt{X_A^2\!+\!b^2}+X_A}\!\right)
\!+\!4M^2\!\left(\!\frac{1}{\sqrt{X_A^2\!+\!b^2}}\!-\!\frac{1}{\sqrt{X_B^2\!+\!b^2}}\!\right)  \right.  \\
\nn&&\hspace*{40pt}\left. +\,\frac{4\,M^2}{b^2}\!\left(X_B-X_A+\sqrt{X_B^2+b^2}-\sqrt{X_A^2+b^2}\,\right)
+\frac{15M^2}{4b}\!\left(\!\arctan{\frac{X_B}{b}}\!-\!\arctan{\frac{X_A}{b}}\!\right) \right.  \\
&&\hspace*{40pt}\left. +\,\frac{M^2}{4}\!\left(\frac{X_A}{X_A^2+b^2}-\frac{X_B}{X_B^2+b^2}\right)\right]~.~~~\label{Integral-8}
\end{eqnarray}

In regard to Eq.~\eqref{Integral-5}, we emphasize two points. First, the analytical 1PM forms for $X_A$ and $X_B$ in Eq.~\eqref{Integral-5} can be determined by the iteration technique as follows:

We adopt an alternative method to determine the integral constant $C$ in Eq.~\eqref{Integral-4} by imposing $t=t_A$ and $x=x_A$. Without loss of generality, we set $t_A=x_A+O(G^2)$ and obtain
\begin{eqnarray}
&&X_A=(1-v)\gamma x_A+O(G^2)~,   \label{XA} \\
\nn&&C=-2(1-v)\gamma M\,ln\left(\sqrt{X_A^2+b^2}+X_A\right)+O(G^2) \\
&&\hspace*{13pt}=-2(1-v)\gamma M\,ln\left[\sqrt{(1-v)^2\gamma^2x_A^2+b^2}+(1-v)\gamma x_A\right]+O(G^2)  ~.   \label{C-1}
\end{eqnarray}

In addition, Eq.~\eqref{Integral-4} up to the 0PM order leads to
\begin{eqnarray}
t_B=x_B+O(G)~.   \label{tB-0}
\end{eqnarray}
Substituting Eqs.~\eqref{C-1} and \eqref{tB-0} into Eq.~\eqref{Integral-4}, up to the 1PM order, we have
\begin{eqnarray}
t_B=x_B+2(1-v)\gamma M\,ln\left[\frac{\sqrt{(1-v)^2\gamma^2x_B^2+b^2}+(1-v)\gamma x_B}{\sqrt{(1-v)^2\gamma^2x_A^2+b^2}+(1-v)\gamma x_A}\right]+O(G^2)~, \label{tB-1}
\end{eqnarray}
where $X_B=\gamma(x_B-v t_B)=(1-v)\gamma x_B+O(G)$ has been used. From Eq.~\eqref{tB-1}, the 1PM form of $X_B$ can be expressed as
\begin{eqnarray}
X_B=(1-v)\gamma x_B-\frac{2vM}{1+v}\,ln\left[\frac{\sqrt{(1-v)^2\gamma^2x_B^2+b^2}+(1-v)\gamma x_B}{\sqrt{(1-v)^2\gamma^2x_A^2+b^2}+(1-v)\gamma x_A}\right] +O(G^2)~.  \label{XB}
\end{eqnarray}

Considering Eqs.~\eqref{XA} and \eqref{XB}, we can express Eq.~\eqref{Integral-5} by the quantities in the background's rest frame $(t,~x,~y,~z)$ as follows
\begin{eqnarray}
\nn&&t\,(B,~A)=(x_B-x_A)+(1-v)\,\gamma \left\{2\,M\,ln\left[\frac{\sqrt{(1-v)^2\,\gamma^2\,x_B^2+b^2}+(1-v)\,\gamma\,x_B}{\sqrt{(1-v)^2\,\gamma^2\,x_A^2+b^2}+(1-v)\,\gamma\,x_A}\right]
+\frac{4M^2}{b^2}\,\times  \right.  \\
\nn&&\hspace*{11pt}\left. \left[\sqrt{\,(1-v)^2\,\gamma^2\,x_B^2+b^2}-\sqrt{\,(1-v)^2\,\gamma^2\,x_A^2+b^2}+(1-v)\,\gamma\,\left(x_B-x_A\right)\right]
+\frac{15M^2-3\,Q^2}{4\,b}\,\times \right. \\
\nn&&\hspace*{11pt}\left. \left[\arctan{\!\frac{(1\!-\!v)\gamma x_B}{b}}\!-\!\arctan{\!\frac{(1\!-\!v)\gamma x_A}{b}}\right]
\!-\!\frac{2\,aM}{b}\!\left[\!\frac{(1-v)\gamma x_B}{\sqrt{(1\!-\!v)^2\gamma^2 x_B^2\!+\!b^2}}
\!-\!\frac{(1-v)\gamma x_A}{\sqrt{(1\!-\!v)^2\gamma^2 x_A^2\!+\!b^2}}\!\right] \right. \\
\nn&&\hspace*{11pt}\left. -\,4\,M^2\!\left[\frac{1}{\sqrt{(1-v)^2\,\gamma^2\,x_B^2+b^2}}-\frac{1}{\sqrt{(1-v)^2\,\gamma^2\,x_A^2+b^2}}\right]
-\frac{M^2-Q^2}{4}\!\left[\frac{(1-v)\,\gamma\,x_B}{(1-v)^2\,\gamma^2\,x_B^2+b^2} \right. \right.  \\
&&\hspace*{11pt}\left. \left. -\frac{(1-v)\,\gamma \,x_A}{(1\!-\!v)^2\gamma^2 x_A^2\!+\!b^2}\right]\!-\!\frac{4\,v\,(1-v)\,\gamma^2M^2}{\sqrt{(1\!-\!v)^2\gamma^2 x_B^2\!+\!b^2}}\,
ln \!\left[\frac{\sqrt{(1\!-\!v)^2\gamma^2 x_B^2+b^2}+(1\!-\!v)\gamma x_B}{\sqrt{(1\!-\!v)^2\gamma^2 x_A^2+b^2}+(1\!-\!v)\gamma x_A}\right] \right\} ~, \label{Integral-9}
\end{eqnarray}
where two series expansions have been performed and the third- and higher-order terms have been dropped. Notice that Eq.~\eqref{Integral-9} is valid for both nonrelativistic and relativistic (such as $v=0.99$) motions of the gravitational source. In the limit $|x_A|\gg b$ and $x_B\gg b$, Eq.~\eqref{Integral-9} is reduced to
\begin{eqnarray}
\nn&&t\,(B,~A)=(x_B\!-\!x_A)+(1\!-\!v)\gamma\!\left\{2M \,ln\!\left[\frac{\sqrt{(1\!-\!v)^2\gamma^2 x_B^2\!+\!b^2}\!+\!(1\!-\!v)\gamma x_B}
{\sqrt{(1\!-\!v)^2\gamma^2 x_A^2\!+\!b^2}\!+\!(1\!-\!v)\gamma x_A}\right]\!+\!\frac{8\,(1\!-\!v)\gamma M^2x_B}{b^2} \right. \\
&&\hspace*{2cm} \left. +\frac{15\,\pi M^2}{4\,b}-\frac{4\,aM}{b}-\frac{3\,\pi\,Q^2}{4\,b} \right\} ~, \label{Integral-10}
\end{eqnarray}
where the second-order terms with the factor $\frac{1}{x_A}$ or $\frac{1}{x_B}$ have been dropped for the same reason mentioned above.

Second, for using Eq.~\eqref{Integral-5}, we can replace the impact parameter $b$ by the coordinates $x_A$ and $y_A$ to express the time delay up to the 2PM order. From Eqs.~\eqref{y} and \eqref{XA}, we obtain the explicit form of $b$ up to the 1PM order by the iteration technique as
\begin{eqnarray}
\nn&&b=-y_A\left(1+\frac{2M}{\sqrt{X_A^2+y_A^2}-X_A}\right)+O(G^2)   \\
&&\hspace*{8pt}=-y_A\left[1+\frac{2M}{\sqrt{(1-v)^2\gamma^2x_A^2+y_A^2}-(1-v)\gamma x_A}\right]+O(G^2)~.   \label{b}
\end{eqnarray}

Plugging Eq.~\eqref{b} into Eq.~\eqref{Integral-9}, up to the 2PM order, we can rewrite Eq.~\eqref{Integral-9} as follow:
\begin{eqnarray}
\nn&&\hspace*{0.5pt} t\,(B,~A)=(x_B-x_A)
+(1-v)\,\gamma \left\{2\,M\,ln\!\left[\frac{\sqrt{(1-v)^2\,\gamma^2\,x_B^2+y_A^2}+(1-v)\,\gamma\,x_B}{\sqrt{(1-v)^2\,\gamma^2\,x_A^2+y_A^2}+(1-v)\,\gamma\,x_A}\right]
\!+\frac{4M^2}{y_A^2}\times  \right.  \\
\nn&&\hspace*{11pt}\left. \left[\sqrt{\,(1-v)^2\,\gamma^2\,x_B^2+y_A^2}-\sqrt{\,(1-v)^2\,\gamma^2\,x_A^2+y_A^2}+(1-v)\,\gamma\,\left(x_B-x_A\right)\right]
\!+\!\frac{15M^2\!-\!3\,Q^2}{4\,y_A}\,\times \right. \\
\nn&&\hspace*{11pt}\left. \left[\arctan{\!\frac{(1\!-\!v)\gamma x_B}{y_A}}\!-\!\arctan{\!\frac{(1\!-\!v)\gamma x_A}{y_A}}\right]
\!+\!\frac{2\,aM}{y_A}\!\!\left[\!\frac{(1-v)\,\gamma\,x_B}{\!\sqrt{(1\!-\!v)^2\gamma^2 x_B^2\!+\!y_A^2}}
\!-\!\frac{(1-v)\,\gamma\,x_A}{\!\sqrt{(1\!-\!v)^2\gamma^2 x_A^2\!+\!y_A^2}}\!\right] \right. \\
\nn&&\hspace*{11pt}\left. -\,4\,M^2\!\left[\frac{1}{\sqrt{(1-v)^2\,\gamma^2\,x_B^2+y_A^2}}-\frac{1}{\sqrt{(1-v)^2\,\gamma^2\,x_A^2+y_A^2}}\right]
\!-\!\frac{M^2\!-\!Q^2}{4}\!\left[\frac{(1-v)\,\gamma\,x_B}{(1-v)^2\,\gamma^2\,x_B^2+y_A^2} \right. \right.  \\
\nn&&\hspace*{11pt}\left. \left. -\,\frac{(1-v)\,\gamma \,x_A}{(1-v)^2\,\gamma^2\,x_A^2+y_A^2}\right]
-\frac{4\,v\,(1-v)\,\gamma^2M^2}{\sqrt{(1-v)^2\,\gamma^2\,x_B^2+y_A^2}}\,
ln\!\left[\frac{\sqrt{(1-v)^2\gamma^2x_B^2+y_A^2}+(1-v)\gamma x_B}{\sqrt{(1-v)^2\gamma^2x_A^2+y_A^2}+(1-v)\gamma x_A}\right]\right.  \\
&&\hspace*{11pt}\left. +\,\frac{4 M^2}{\sqrt{(1\!-\!v)^2\gamma^2 x_B^2+y_A^2}}\frac{\sqrt{(1\!-\!v)^2\gamma^2 x_B^2+y_A^2}-(1\!-\!v)\gamma x_B}
{\sqrt{(1\!-\!v)^2\gamma^2 x_A^2+y_A^2}-(1\!-\!v)\gamma x_A}-\frac{4M^2}{\sqrt{(1\!-\!v)^2\gamma^2 x_A^2+y_A^2}}\right\}  ~.\label{Integral-11}
\end{eqnarray}
Correspondingly, for the case of $|x_A|\gg |y_A|$ and $x_B\gg |y_A|$, Eq.~\eqref{Integral-11} can be simplified to
\begin{eqnarray}
\nn&&t\,(B,~A)=(x_B-x_A)+(1\!-\!v)\gamma \left\{2\,M\,ln\left[-\frac{4\,(1\!-\!v)^2\gamma^2\,x_A\,x_B}{y_A^2} \right]
\!+\!\frac{8\,(1\!-\!v)\gamma M^2x_B}{y_A^2}\!-\!\frac{15\,\pi M^2}{4\,y_A} \right. \\
&&\hspace*{2cm} \left. +\frac{4\,aM}{y_A}+\frac{3\,\pi\,Q^2}{4\,y_A} \right\} ~,\label{Integral-12}
\end{eqnarray}
where the the second-order terms with the factor $\frac{1}{x_A}$ or $\frac{1}{x_B}$ have been ignored.

\section{Magnitudes of the kinematically correctional effects on the delay} \label{Magnitude}
In this section we analyze the magnitudes of the velocity effects on the time delay. We consider nonrelativistic as well as relativistic cases for the motion of the gravitational source, since there are some celestial bodies moving with a high radial velocity~\cite{Ghezetal2008,MMGWMLGR2011,Zhengetal2014}. For illustration, we take Eq.~\eqref{Integral-12} as an example. In order to evaluate their magnitudes, we follow the notations in Ref.~\cite{LinHe2016b} and present the general forms of the velocity-induced correctional effects in Eq.~\eqref{Integral-12} as follows
\begin{eqnarray}
&&\Delta_{FM}(v)= 2M\left\{ ln \!\left(-\frac{4\,x_A\,x_B}{y_A^2}\right)-\sqrt{\frac{1-v}{1+v}}\,ln\!\left[-\frac{4\,(1-v)\,x_A\,x_B}{(1+v)\,y_A^2} \right]\right\}~,  \label{first-M} \\
&&\Delta_{SM-1}(v)=\frac{16\,v\,M^2\,x_B}{(1+v)\,y_A^2}~,   \label{second-M-1}   \\
&&\Delta_{SM-2}(v)=-\left(\,1-\sqrt{\frac{1-v}{1+v}}\,\right)\frac{15\,\pi\,M^2}{4\,y_A}~,         \label{second-M-2}   \\
&&\Delta_{a}(v)=\left(\,1-\sqrt{\frac{1-v}{1+v}}\,\right)\frac{4\,a\,M}{y_A}~,                     \label{second-a}     \\
&&\Delta_Q(v)=\left(\,1-\sqrt{\frac{1-v}{1+v}}\,\right)\frac{3\,\pi\,Q^2}{4\,y_A}~,                \label{second-Q}
\end{eqnarray}
where $\Delta_{FM}(v)$, $\Delta_{a}(v)$, and $\Delta_Q(v)$ denote the velocity corrections to the first-order Schwarzschild, second-order Kerr, and charge-induced terms on the right-hand side of Eq.~\eqref{Integral-7}, respectively. $\Delta_{SM-1}(v)$ and $\Delta_{SM-2}(v)$ represent the velocity corrections to the larger and smaller second-order Schwarzschild contributions to the delay in Eq.~\eqref{Integral-7}, respectively.

\begin{table}[t]
\begin{minipage}[b]{13cm}
  \begin{tabular}{cccccccc}
        \hline
                      $v$       & $~~~\Delta_{FM}(v)~~~$ &~~~$\Delta_{SM-1}(v)~~~$ & $~~~\Delta_{SM-2}(v)~~~$ & $~~~\Delta_a(v)~~~$ & $~~~\Delta_Q(v)~~~$     \\
       \hline
                    0.9         &       609$\,\mu s$     &      0.186$\,\mu s$     &        0.223$\,ns$       &   $-$\,7.58$\,ps$   &   $-$\,4.46$\,fs$       \\
                    0.1         &       80.3$\,\mu s$    &      35.8$\,ns$         &        27.6$\,ps$        &   $-$\,0.939$\,ps$  &       $\star$           \\
                  0.001         &       0.845$\,\mu s$   &      0.393$\,ns$        &        0.289$\,ps$       &   $-$\,9.83$\,fs$   &       $\star$           \\
                0.00001         &       8.46$\,ns$       &      3.93$\,ps$         &        2.90$\,fs$        &        $\star$      &       $\star$           \\
              0.0000001         &       84.6$\,ps$       &      39.3$\,fs$         &         $\star$          &        $\star$      &       $\star$           \\
       \hline
  \end{tabular}\par
\end{minipage}
\caption{The magnitudes of the kinematical corrections to the light delay for various $v$, with the lens's mass $M$ being $5M_{\odot}$. Hereafter, our attention is concentrated on the absolute values $(\geq0)$ of these magnitudes, and the star ``$\star$" denotes the absolute value which is less than $1\,fs$. We present the case with non-zero electrical charge mainly for illustration. }   \label{Table1}
\end{table}
\begin{table}[t]
\begin{minipage}[b]{13cm}
  \begin{tabular}{cccccccc}
        \hline
                      $v$       & $~~~\Delta_{FM}(v)~~~$ &~~~$\Delta_{SM-1}(v)~~~$ & $~~~\Delta_{SM-2}(v)~~~$ & $~~~\Delta_a(v)~~~$ & $~~~\Delta_Q(v)~~~$     \\
       \hline
                    0.9         &       0.122$\,s$       &      37.3$\,\mu s$      &        44.6$\,ns$        &   $-$\,1.52$\,ns$   &   $-$\,0.893$\,ps$      \\
                    0.1         &       0.0161$\,s$      &      7.15$\,\mu s$      &        5.53$\,ns$        &   $-$\,0.188$\,ns$  &   $-$\,0.111$\,ps$      \\
                  0.001         &       169$\,\mu s$     &      78.6$\,ns$         &        57.9$\,ps$        &   $-$\,1.97$\,ps$   &   $-$\,1.16$\,fs$       \\
                0.00001         &       1.69$\,\mu s$    &      0.787$\,ns$        &        0.580$\,ps$       &   $-$\,19.7$\,fs$   &       $\star$           \\
              0.0000001         &       16.9$\,ns$       &      7.87$\,ps$         &        5.79$\,fs$        &        $\star$      &       $\star$           \\
       \hline
  \end{tabular}\par
\end{minipage}
\caption{The magnitudes of the kinematical corrections for the lens with an intermediate mass $M=1000M_{\odot}$. }   \label{Table2}
\end{table}
As an example, the related parameters are preset as follows. We set $y_A=-1.0\times10^{6}M$ to guarantee a weak field. $x_B$ and $|x_A|$ are much larger than $|y_A|$ and set to be $x_B=-x_A=1.0\times10^{9}M~(=-1000\,y_A)$, $a=0.1M$, and $Q=0.01M$. Notice that for certain parameters $y_A$, $x_A$, $x_B$, $a$, and $Q$, the correctional effects defined in Eqs.~\eqref{first-M} - \eqref{second-Q} are not only dependent on the radial velocity $v$ but also proportional to the mass $M$ of the gravitational source. Moreover, it is generally believed that there are three classifications for black holes by their masses in our Universe, i.e., stellar-mass ($M\!\sim\! 3-20M_{\odot}$), intermediate-mass ($M\!\sim\!100-10^4M_{\odot}$), and supermassive ($M\sim10^{6}-10^{10}M_{\odot}$) black holes~\cite{MC2004,NM2013}. Therefore, we typically assume that the rest mass $M$ of the moving KN black hole to be $5,~1000$, and $4.0\times10^{6}M_{\odot}$ in Tabs.~\ref{Table1} - \ref{Table3}, respectively, to show the magnitudes of these correctional effects for various velocities of the lens, with $M_{\odot}~(=1.475km)$ being the mass of the Sun.

\begin{table}[t]
\begin{minipage}[b]{13cm}
  \begin{tabular}{cccccccc}
        \hline
                      $v$       & $~~~\Delta_{FM}(v)~~~$ &~~~$\Delta_{SM-1}(v)~~~$ & $~~~\Delta_{SM-2}(v)~~~$ & $~~~\Delta_a(v)~~~$ & $~~~\Delta_Q(v)~~~$     \\
       \hline
                    0.9         &       487$\,s$       &       0.149$\,s$          &       179$\,\mu s$       & $-$\,6.06$\,\mu s$  &   $-$\,3.57$\,ns$       \\
                    0.1         &       64.2$\,s$      &       0.0286$\,s$         &       22.1$\,\mu s$      & $-$\,0.751$\,\mu s$ &   $-$\,0.442$\,ns$      \\
                  0.001         &       0.676$\,s$     &       314$\,\mu s$        &       0.232$\,\mu s$     & $-$\,7.86$\,ns$     &   $-$\,4.63$\,ps$       \\
                0.00001         &       0.00677$\,s$   &       3.15$\,\mu s$       &       2.32$\,ns$         & $-$\,78.7$\,ps$     &   $-$\,46.3$\,fs$       \\
              0.0000001         &       67.7$\,\mu s$  &       31.5$\,ns$          &       23.2$\,ps$         & $-$\,0.787$\,ps$    &       $\star$           \\
       \hline
  \end{tabular}\par
\end{minipage}
\caption{The magnitudes of the kinematical corrections for the lens with a supermassive mass $M=4.0\times10^{6}M_{\odot}$. }   \label{Table3}
\end{table}

\section{Possible applications and detection of the velocity effects on the second-order delay} \label{detection}

Velocity effects on the second-order gravitational delay may influence on the high-precision measurements of some crucial parameters, such as the post-Newtonian parameters~\cite{KopeiFoma2007,BAI2008,Kopeikin2009}, and Hubble's constant when measured via the time delay between two lensed images~\cite{Refsdal1964,LJRWKTME1998,FXKR2002,GJKFB2014}. For example, the relation between the time delay and the post-Newtonian parameters in the time-dependent gravitational field has been given in Ref.~\cite{Kopeikin2009} (see Eq.~(71) therein). The velocity effects on the static time delay make the observed numerical values of these post-Newtonian parameters biased. Therefore, it is worthwhile to consider their possible detection.

Based on the results shown in Tabs.~\ref{Table1} - \ref{Table3}, we analyze the possibility of detecting the velocity effects qualitatively.
For the convenience of our discussion, we fix the parameters $x_A$, $x_B$, and $y_A$ as given above. We focus on the velocity effects on the second-order contributions to the delay, since the correctional effect on the first-order delay has been studied in detail.

We first consider $\Delta_{SM-1}(v)$. For a supermassive black hole with mass $M=4.0\times10^{6}M_{\odot}$, Tab.~\ref{Table3} shows that the kinematically correctional effect ($\,> 1 ns$) on the larger second-order Schwarzschild contribution to the delay is much larger than the accuracy ($\sim ps$) of today's high-precision techniques, for both relativistic and nonrelativistic motions of the black hole. For example, $\Delta_{SM-1}(v)$ is about $31.5\,ns$ when this black hole moves at an extremely low radial velocity $v=30\,m/s$. Thus, the possibility of detecting the correctional effect $\Delta_{SM-1}(v)$ might be very large for moving supermassive black holes. This conclusion also holds for moving black holes with an intermediate mass, since the magnitude of $\Delta_{SM-1}(v)$ in the case of $M=1000M_{\odot}$ is larger than $1ps$ for almost all of the range of the radial velocity $v$, as presented in Tab.~\ref{Table2}. Even for a stellar-mass black hole moving with a low radial velocity, we find there exists the possibility to detect $\Delta_{SM-1}(v)$. For instance, $\Delta_{SM-1}(v)$ still exceeds $1ps$ for the case of $v=3km/s$ in Tab.~\ref{Table1}.

We then discuss the correctional effect on the smaller second-order Schwarzschild contribution to the delay, i.e., $\Delta_{SM-2}(v)$. From Tab.~\ref{Table3}, we notice that $\Delta_{SM-2}(v)$ can be larger than $1ps$ for a supermassive black hole $(M=4.0\times10^{6}M_{\odot})$ moving at a very low velocity. Thus, it is still very likely to detect $\Delta_{SM-2}(v)$ for moving supermassive black holes. When the mass of the lens decreases from the supermassive class to the stellar class, the motion of the gravitational source has to change from a nonrelativistic case to a relativistic case, for the possible detection of $\Delta_{SM-2}(v)$.

Rotating black holes such as the Sagittarius A$^*$ (a supermassive black hole) in the Galactic center~\cite{ZLY2015} are very common in our universe. It is also necessary to take the velocity effect on the Kerr delay into account. Tabs.~\ref{Table1} and \ref{Table2} indicate that it is possible to detect $\Delta_a(v)$ only when the lens' motion tends to be relativistic for stellar-mass and intermediate-mass black holes. In contrast to these relatively small black holes, we might observe $\Delta_a(v)$ for supermassive black holes with a nonrelativistic radial velocity. For example, $\Delta_a(v)$ can largely exceed $1ps$ for the case of $M=4.0\times10^{6}M_{\odot}$ and $v=3km/s$.

With respect to the correctional effect on the charge-induced contribution to the delay, namely $\Delta_Q(v)$, we have to conclude that there may not be any chance for its possible detection via today's techniques. As shown in Tab.~\ref{Table3}, even for a supermassive black hole with $M=4.0\times10^{6}M_{\odot}$ and $v=300km/s$, $\Delta_Q(v)$ is still a little larger than $1ps$. In addition, the original charge of a black hole in the Universe might have been neutralized or become very small in most cases, which makes the detection more difficult. However, recently, some new techniques to detect tiny time delays of light in the framework of special relativity have been proposed~\cite{BS2010,SB2013}, with an unprecedented ultimate precision limit which is much less than $1\,fs$. There is the possibility that they are developed into the astronomical domain for the measurements of Shapiro delays, and $\Delta_Q(v)$ might also be observed at that time.

\section{Summary} \label{Summary}
In this work we calculate the second-order gravitational time delay of light propagating in the equatorial plane and parallel to the velocity of a constantly moving Kerr-Newman black hole, based on the 2PM harmonic metric. With respect to the velocity effects, we find that the relativistic correctional factor $(1-v)\gamma$ applies not only to the first-order term but also to all the second-order terms in the time delay. We also analyze the magnitudes of the correctional effects on the second-order contributions to the delay and their possible detections. We conclude that it is likely to detect the velocity effects on the second-order Schwarzschild and Kerr contributions to the delay by today's high-accuracy techniques such as the VLBI. Our result and conclusions might be useful in future astronomical observations.

\section*{ACKNOWLEDGEMENTS}
We would like to thank the referee very much for valuable comments and suggestions to improve the quality of this work. This work was supported in part by the National Natural Science Foundation of China (Grant No. 11547311), the National Basic Research Program of China (Grant No. 2013CB328904), and the Fundamental Research Funds for the Central Universities (Grant No. 2682014ZT32).

\end{document}